# Spaser Spectroscopy with Subwavelength Spatial Resolution

*Yuriy E. Lozovik[†‡], Igor A. Nechepurenko[‡§], Alexander V. Dorofeenko[‡§], Alexander A. Pukhov[‡§], Eugeniy S. Andrianov[‡§]*

[†]Institute of Spectroscopy RAS, Moscow Region, Troitsk, Russia

[‡]Moscow Institute of Physics and Technology, Moscow Region, Dolgoprudny, Russia

[§]Institute for Theoretical and Applied Electromagnetics RAS, Moscow, Russia

Keywords: spaser; laser spectroscopy; surface plasmons; SNOM

A new method for high-sensitivity subwavelength spectromicroscopy is proposed based on the usage of a spaser (near-field laser) in the form of a scanning probe microscope tip. The high spatial resolution is defined by the tip's curvature, as is the case for apertureless scanning near-field optical microscopy. In contrast to the latter method, we suggest using radiationless plasmon pumping by neighbouring quantum dots instead of irradiation of the tip by an external laser beam. The spaser generation spectrum is analyzed. The plasmon generation is suppressed due to absorption at the transition frequencies of the neighbouring nano-objects (molecules or clusters) under study. As a result, narrow dips appear in the wide plasmon generation spectrum. Further, the highest sensitivity is achieved near the spaser generation threshold. The sensitivity of the spaser spectromicroscope is estimated.

Recent developments in electrodynamics have largely been governed by a tendency to expand the application range of optical devices to fields where electronic, magnetic, X-ray and other devices were conventionally used. Some examples are future optical computers,[1,2] optical memory,[3] optical lithography of high (subwavelength) spatial resolution,[4,5] as well as scanning near-field optical microscopy (SNOM).[6-15] It is the use of the optical near-field that makes it possible to overcome the Rayleigh limit of spatial resolution $\sim \lambda$, which determines the lower bound for the size of optical devices. The aperture realization of SNOM is based on light transmission through the small (subwavelength) hole. Since this is a tunneling process, it results in a small analyzed signal intensity and thus decreases sensitivity. This shortcoming can be overcome in apertureless schemes including metal, where the incident electromagnetic wave is enhanced near a narrow needle tip of a scanning probe microscope (SPM) due to the "lighting rod effect" and resonance plasmon excitation at the tip. The greatest enhancement is achieved when the near-field excitations, surface plasmon-polaritons (SPPs),[16] are excited at the tip. SPPs are capable of creating a relatively large field intensity in a small (subwavelength) spatial region,[17] which is used in plasmonic lenses.[18-22] This property is also used in apertureless SNOM, where the near-field (in the region near the tip of the SPM) created by the incident laser beam, is scattered by an analyzed sample. This creates a signal in the far field, whose intensity (and in some schemes, phase) is measured and used for image retrieval. SNOM enables high-resolution images of the surface profile to be obtained, but the information about the sample composition is not provided by this method.

The latter limitation is overcome in Tip-Enhanced Optical Spectroscopy (TEOS), which exploits the principles of SNOM but analyzes the spectral response of the sample.[23] Thus, both the image and the composition of the sample are obtained. The most popular is the method based on Raman spectroscopy, Tip-Enhanced Raman Spectroscopy (TERS),[24] which takes advantage of the large field intensity in SPPs for an even larger enhancement of the nonlinear effect.

However, the most sensitive spectral method is intracavity laser spectroscopy[25-27] which offers no spatial resolution, but provides an extremely high sensitivity and reveals even forbidden (non-dipolar) transitions.[28] This sensitivity is related to multipass (restricted by the cavity quality) of light through the sample. We believe that further development of spectroscopy should combine the TEOS and intracavity laser spectroscopy methods in order to combine their benefits.

In the current paper, we suggest a novel method of spectromicroscopy with ultrahigh spatial resolution and high sensitivity. These properties are inherited from the TEOS and laser spectroscopy methods, although the proposed technique is very different from both of them and should lead to the appearance of a new class of methods for the study of nano-objects and surfaces. The new method uses a spaser, a quantum plasmonic device which generates plasmons due to nonradiative energy transfer from the gain medium (a quantum dot) to SPPs, localized at a nanoparticle,[29-34] or travelling along a 1D metallic object.[35] The device suggested here is based on a 1D spaser generating plasmons on a needle with a narrow tip. The analyzed nano-object interacts effectively with the field of the plasmonic mode, concentrated near the needle tip[36-40] and can resonantly absorb the plasmon quanta. This process is analogous to the interaction with the field inside a laser cavity used in intracavity laser spectroscopy. The plasmonic lasing (spasing) can give a much higher field intensity than the TEOS scheme since the plasmon at the tip is excited not by the scattering of the incident wave, but by direct nonradiative energy transfer from quantum dots or from another gain medium placed on the needle. The high field intensity favours high sensitivity even for linear methods, to say nothing of TERS. Additionally, in the case of current pumping, the sample is not exposed to any external radiation. The advantage of the proposed method over standard laser spectroscopy, besides the high spatial resolution, is the weakness of the nano-sized spaser compared to the macroscopic laser. In standard intracavity laser spectroscopy, the highest sensitivity is observed near the threshold,[41] where the lasing becomes "weak". In our case, the oscillation of the nano-sized laser is easily suppressed by absorption in the analyzed sample, which gives outstanding sensitivity.

A principal scheme of the spaser spectroscope is shown in Fig. 1. The needle geometry typical of near-field devices supports a plasmonic solution. A mode propagating along the needle should be localized between the needle tip and a Bragg reflector,[42] the latter realized as a periodic corrugation of the needle surface. Thus, a cavity is formed for a plasmonic mode, which is amplified by a set of quantum dots deposited onto the needle or by any other gain medium; a 1D spaser is formed. The field enhanced by the tip (Fig. 2) interacts effectively with the analyzed sample. The field localization ensures spatial resolution of the order of the tip's curvature radius. The wave, having tunneled through the Bragg reflector, is transferred to a spectral analyzer and may be amplified by additional gain.

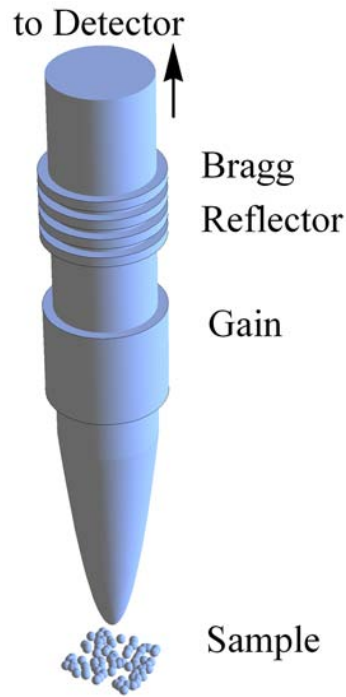

Fig. 1. Principal scheme of the tip based spaser spectroscopy device.

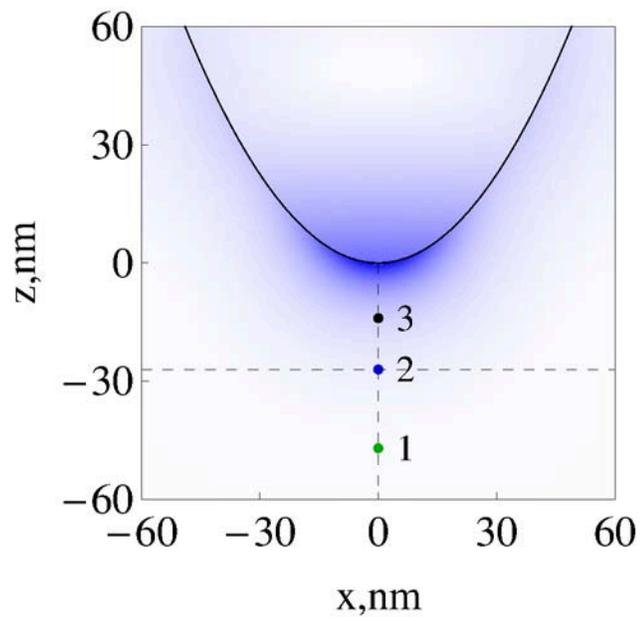

Fig. 2. The electric field distribution near the tip. The crossing of the dashed lines (point 2) denotes the point of optimum sample position.

In the classical methods of laser spectroscopy,[25-27] the spectral line of the analyzed sample is scanned by tuning the frequency of the laser with very narrow line, or by the use of a multimode laser.[28] In the scheme based on the spaser, the line width is increased by a few orders of magnitude due to strong noise in the metal. Indeed, since the cavity mode has a lifetime smaller than the transverse relaxation time $T_2$ of the quantum dots, the spasing line width is equal to $T_2^{-1}$. The absorption peaks of the sample molecules are often narrower than the spasing line. Such a situation is not characteristic of traditional lasers, and it is not obvious whether a narrow absorption peak suppresses the whole spasing, or creates a dip in the spectral line. It is shown below that each spectral line of the sample is displayed as a dip in the spectral line of the spaser. Therefore, the device suggested here has no tunable elements in its simplest realization, although they can be included to expand the working range.

In most papers devoted to traditional laser spectroscopy, one reasonably neglects quantum noise effects,[28] except when working near the threshold.[41] The ordinary situation is a set of dense and narrow cavity modes, some of which are suppressed by absorption in the sample particles. Alternatively, one uses a single-mode laser with a very narrow spectral line. Then the absorption spectrum is scanned by tuning the cavity frequency. In all these cases, consideration of the quantum noise is generally of no use. In our case, we use an approximation of single mode plasmonic lasing (spasing), homogeneously broadened by noise in metallic needle. First, the single-mode approximation is justified by the possibility of selecting a plasmonic mode of the needle. Although a metallic cylinder has a set of plasmonic solutions with different dispersion relations $\omega_i(k)$, $i = 0,1,2,...$, the frequency $\omega$ is specified by the transition of quantum dots, and the wave number is determined by the distance between the needle tip and the Bragg reflector (see Fig. 1). Second, the consideration of broadening is necessary because the sample spectrum is observed within the spaser's spectral line. Its width is completely due to noise. According to the fluctuation-dissipation theorem, the noise source is also the source of dissipation in metal.

The spaser dynamics is described by Maxwell-Bloch equations, which follow from the Maxwell equations and from the equations for the density matrix of the gain medium atoms. These are partial differential equations, which are reduced to ordinary differential equations by use of the single-mode approximation,[43] which represents the electric field as a product of a (dimentionless) time amplitude and a cavity mode $\mathbf{E}(\mathbf{r})$. In our case, the latter is the field distribution in the plasmonic mode of the needle, restricted by the tip and the Bragg reflector. Further simplification of the equations is attained by using the slow amplitudes approximation, which reduces the second-order equations to first-order ones. This gives a system of equations for the amplitude of the cavity mode $a$, polarization of the gain medium $\sigma$ and population inversion $D$:[44]

$$\dot{a} = -a/\tau_a - i\Omega\sigma + F(t), \tag{1a}$$

$$\dot{\sigma} = (i\delta - 1/T_2)\sigma + i\Omega aD, \tag{1b}$$

$$\dot{D} = -(D - D_0)/T_1 + 2i\Omega(a^*\sigma - \sigma^*a). \tag{1c}$$

These equations show that the population inversion is pushed towards the value of $D_0$ by pumping and is simultaneously suppressed by field. At positive values of $D$, oscillations of $\sigma$ and $a$ are produced; as a result, spasing appears.

The system's behaviour is modeled at realistic parameters. The longitudinal and transverse relaxation times are $T_1 \sim 10^{-13}$ s and $T_2 \sim 10^{-13}$ s for CdSe quantum dots. The cavity mode (plasmon) relaxation time is $\tau_a = (2\pi)^{-1}\left(\int \partial(\varepsilon'\omega)/\partial\omega \mathbf{EE}^* dV\right)/\left(\int \varepsilon''\mathbf{EE}^* dV\right) \sim 10^{-14}$ s. Here, the plasmon field distribution $\mathbf{E}(\mathbf{r})$ is assessed for the infinite cylinder of permittivity characteristic of silver at $\lambda = 350$ nm.[45] The interaction constant is given as $\Omega = \sqrt{\mu\omega/8\hbar W}$, where the mode energy is $W = (8\pi)^{-1}\int \partial(\varepsilon'\omega)/\partial\omega \mathbf{EE}^* dV$, and the confinement factor[43] may be approximated as $\mu \sim |\mathbf{d}_{12}|^2 \langle \mathbf{E}^2 \rangle_\Sigma \sim 10^{12} \div 10^{13}$ s (the subscript "$\Sigma$" denotes an average over the metal surface, where the quantum dots are located). This gives the evaluation $\Omega \sim 10^{12} \div 10^{13}$ $s^{-1}$. Detuning $\delta = \omega_a - \omega_0$ of the cavity frequency from the gain medium transition frequency is taken to be zero.

Without taking account of the noise $F(t)$, the stationary lasing spectrum would be a Dirac delta function. The simplest approach to the spectrum calculation is a direct solution of the Langevin-type equation system (1) using the delta-correlated (white) noise, $\langle F(t)F(t+\tau)\rangle = 2\delta(\tau)/\tau_a$, where the angle brackets correspond to an average over the random process realizations, and the factor $2/\tau_a$ follows from the analysis of the system's interaction with the reservoir. This interaction leads to both cavity mode dissipation and fluctuations.[46] The direct calculation of the spectrum of the stochastic signal $a(t)$ is challenging because the signal is unbounded in time. On the other hand, the autocorrelator $C(t) = \langle a^*(0)a(t)\rangle$ decays exponentially when time goes to infinity.[47] Therefore, a convenient way to approach the lasing spectrum calculation is to use the Wiener-Khinchin theorem $S(\omega) = (2\pi)^{-1}\int_{-\infty}^{\infty} C(t)\exp(i\omega t)dt$. This result is shown by the unnumbered line in Fig. 3.

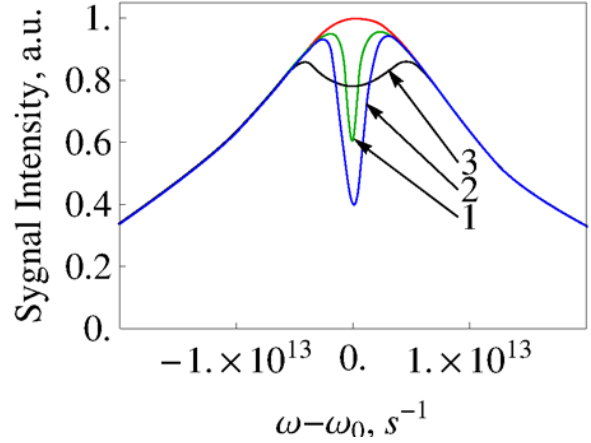

Fig. 3. The lasing spectrum of the spaser (unnumbered line) and its change by interaction with the sample (the curve numbers correspond to points in Fig. 2).

Interaction of the plasmonic laser with nano-objects, which resonantly absorb plasmon quanta in some narrow spectral range, leads to the appearance of a dip in the spasing spectrum. The absorbing nano-objects are described by equations similar to (1b), (1c) with negative pumping. Such an approach takes absorber saturation effects into account. Thus, the system (1) assumes the following form:

$$\dot{a} = -a/\tau_a - i\Omega\sigma - i\Omega_S\sigma_S + F(t),$$

$$\dot{\sigma} = (i\delta - 1/T_2)\sigma + i\Omega aD, \qquad (2)$$

$$\dot{D} = -(D - D_0)/T_1 + 2i\Omega(a^*\sigma - \sigma^* a),$$

$$\dot{\sigma}_S = (i\delta_S - 1/T_{2S})\sigma_S + i\Omega_S aD_S,$$

$$\dot{D}_S = -(D_S - D_{0S})/T_{1S} + 2i\Omega_S(a^*\sigma_S - \sigma_S^* a).$$

The subscript "S" denotes the variables related to the analyzed sample. The value of $D_{0S}$ is proportional to the number of sample absorbing particles (atoms, molecules, etc.), $D_{0S} = N\tilde{D}_{0S}$. The value of $\tilde{D}_{0S}$ is the difference between the populations of the top and bottom levels for a single particle and should be negative to describe absorption rather than gain, $-1 \leq \tilde{D}_{0S} < 0$. In our calculations, we assume that in the absence of a field, the particles are in their ground state, i.e., $\tilde{D}_{0S} = -1$. The longitudinal and transverse relaxation times for the sample particles may vary over a wide range; we have used the values $T_{1S} \sim 4 \cdot 10^{-12}$ s and $T_{2S} \sim 4 \cdot 10^{-12}$ s. The effect of the sample, positioned near the tip, is determined by its overlap with the plasmon field: $\Omega_S = \sqrt{\mu_S \omega / 8\hbar W}$ and $\mu_S \sim |\mathbf{d}_{12S}|^2 \langle \mathbf{E}^2 \rangle_S$ where $\langle ... \rangle_S$ denotes an average over the sample. The field distribution (Fig. 2) was calculated for the parabolic form of the tip. Since its curvature radius $\rho \sim 20$ nm is far below the wavelength, the field was found in the electrostatic approximation using parabolic coordinates. In fact, the value of $\Omega_S$ is tuned by a vertical shift of the needle (Fig. 4a) and may be considered to be a free parameter.

The numerical solution of the equation system (2) gives the spasing spectrum when the Wiener-Khinchin theorem is applied. For a weak interaction (where the needle is far from the sample, i.e., at small values of $\Omega_S$), the dip in the generation spectrum is also weak (curve 1 in Fig. 3). Growth of the interaction increases the dip (curve 2 in Fig. 3), but finally broadens it (curve 3 in Fig. 3) which leads to a decrease of its depth and of the spectral resolution. Thus, an optimal needle position exists which may be found from the maximum dip depth (Fig. 5). Alternatively, one may use the "quality factor" of the dip.

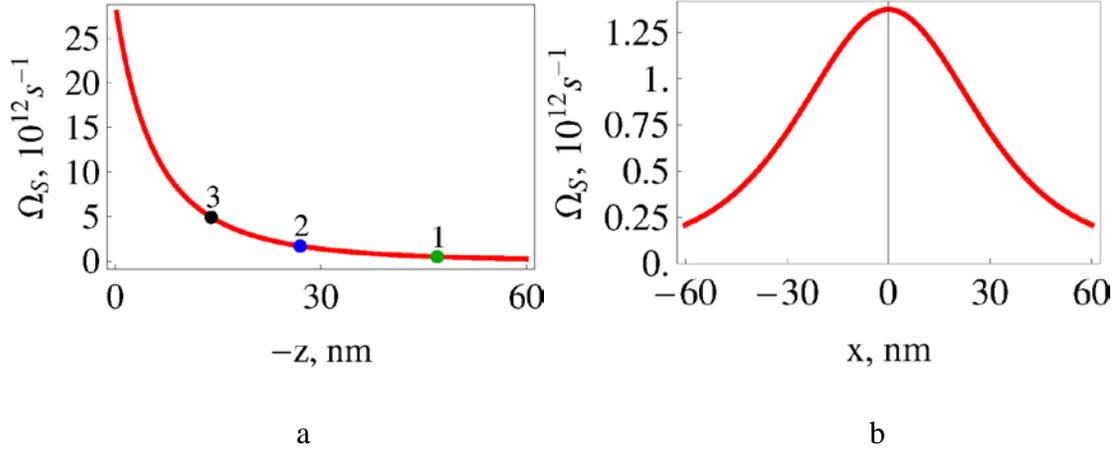

Fig. 4. Interaction constant of the spaser with an analyzed object as a function of vertical (a) and horizontal (b) coordinates ($D_0 = -20$, see the text). The point numbers correspond to those in Fig. 2.

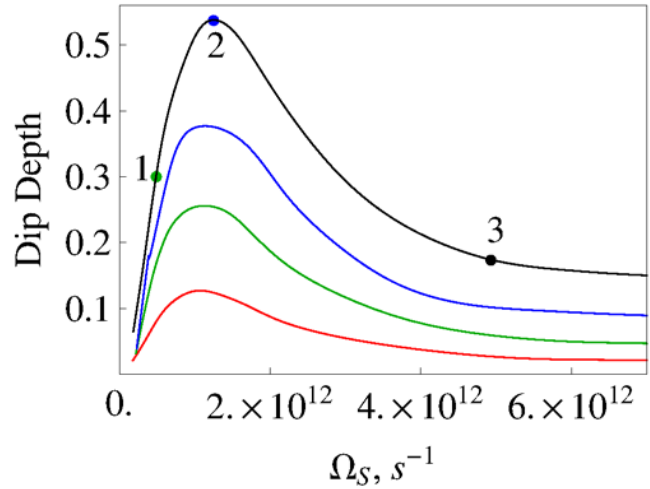

Fig. 5. Dependence of the spectral dip depth on the interaction constant $\Omega_S$ (i.e., on the vertical position of the needle). The numbers correspond to those in Figs. 2–4. The curves (from the bottom to the top) correspond to 5, 10, 15 and 20 sample atoms, respectively.

The calculations demonstrate that the device easily detects some dozens of atoms. The spatial resolution is illustrated in Fig. 4b. It is seen that the spot around the optimal point, where the interaction constant is substantial, is of the order of the tip curvature radius (20 nm). The influence of the atoms appearing to the side of the optimum point may be estimated from Fig. 5 since the value of $\Omega_S$ describes the interaction irrespective of the sample shift direction. Thus, the following curious feature appears. If the tip is too close to the analyzed surface, then the region, which creates the best resolved spectrum (similar to curve 2 in Fig. 3), forms a circle. Inside this circle, a "broadened-spectrum region" is located where the spectrum is similar to that shown by the curve 3 in Fig. 3, whereas the outside region is weakly interacting (the spectrum is similar to that shown by curve 1 in Fig. 3). Presumably, such a response may cause confusion, and it is better to place the tip at the optimal or at a slightly larger distance from the sample.

Let us note that the possibility of the detection of tiny quantities of a substance follows from the nanoscopic size of the spaser. Indeed, to create a dip in the spaser spectrum, the absorbing particles should suppress oscillations at their transition frequency. Of course, such suppression is much easier in the case of a spaser than in the case of traditional (macroscopic) laser, which is responsible for a high sensitivity of spaser spectroscopy.

It should be also noted that in the present paper, the spasing line broadening was modeled through the noise in metal. Another possible mechanism, namely the inhomogeneous broadening of the gain medium transition line, will be considered in the future.

In conclusion, in this paper, a novel method of scanning spaser spectromicroscopy is suggested. It combines the benefits of tip-enhanced optical spectroscopy and of intracavity laser spectroscopy. The use of a nanolaser (spaser) as a source of plasmons in the TEOS scheme increases the field intensity. On the one hand, in comparison with traditional laser spectroscopy, a nano-sized spaser is damped easily by only a few dozen sample atoms. On the other hand, the spaser has a very broad spectral line, and instead of scanning over the sample spectrum, one can observe the absorption lines as dips in the spasing line.

AUTHOR INFORMATION

**Corresponding Author**

*Email: alexandor7@gmail.com

ACKNOWLEDGMENTS

The authors thank Prof. Alexey Vinogradov for useful discussions. The work is partly supported by the RFBR grants 10-02-00857-a, 10-02-91750-AF_a, 11-02-92475-MNTI_a, 12-02-01093-a and by the Dynasty Foundation.